\documentclass[twocolumn,
groupedaddress,
superscriptaddress,
amsmath,amssymb,
aps,
pra,
]{revtex4-1}
\usepackage{amsmath}
\usepackage{subfigure}
\usepackage{graphicx}
\usepackage{hyperref}
\hypersetup{
	colorlinks = true, linkcolor = blue, anchorcolor = blue, urlcolor = blue, citecolor = blue
}
\usepackage{xcolor}

\usepackage{verbatim}

\begin{document}
	
	
	\title{Scalable high-rate, high-dimensional time-bin encoding quantum key distribution}
	
	
	\author{Nurul T. Islam}
	\email[]{nurul.taimur.islam@gmail.com}
	\affiliation{Department of Physics, The Ohio State University, 191 West Woodruff Ave., Columbus, Ohio 43210 USA}
	\affiliation{Department of Physics and the Fitzpatrick Institute for Photonics, Duke University, Durham, North Carolina 27708, USA}
	\author{Charles Ci Wen Lim}
	\email[]{charles.lim@nus.edu.sg}
	\affiliation{Department of Electrical and Computer Engineering, National University of Singapore, 117583, Singapore}
    \affiliation{Centre for Quantum Technologies, National University of Singapore, 117543, Singapore}
	\author{Clinton Cahall}
	\affiliation{Department of Electrical Engineering and the Fitzpatrick Institute for Photonics, Duke University, Durham, North Carolina 27708, USA}
	\author{Bing Qi}
	\affiliation{Quantum Information Science Group, Computational Sciences and Engineering Division,
	Oak Ridge National Laboratory, Oak Ridge, TN 37831-6418, USA}
	\author{Jungsang Kim}
	\affiliation{Department of Electrical Engineering and the Fitzpatrick Institute for Photonics, Duke University, Durham, North Carolina 27708, USA}
    \affiliation{IonQ, Inc., College Park, MD 20740, USA}
	\author{Daniel J. Gauthier}
	\affiliation{Department of Physics, The Ohio State University, 191 West Woodruff Ave., Columbus, Ohio 43210 USA}


\date{\today}

\begin{abstract}

We propose and experimentally demonstrate a new scheme for measuring high-dimensional phase states using a two-photon interference technique, which we refer to as quantum-controlled measurement. Using this scheme, we implement a $d$-dimensional time-phase quantum key distribution (QKD) system and achieve secret key rates of 5.26 and 8.65 Mbps using $d = 2$ and $d = 8$ quantum states, respectively, for a 4~dB channel loss, illustrating that high-dimensional time-phase QKD protocols are advantageous for low-loss quantum channels. This work paves the way for practical high-dimensional QKD protocols for metropolitan-scale systems. Furthermore, our results apply equally well for other high-dimensional protocols, such as those using the spatial degree-of-freedom with orbital angular momentum states being one example. 

\end{abstract}


\maketitle
\section{Introduction}
Interference of two photons is central to many important quantum information technologies, such as linear optics-based quantum computing~\cite{Makinoe1501772}, quantum metrology~\cite{Lyonseaap9416} and sensing~\cite{Dowling_QuantumMetrology}. Recently, two-photon interference has been used in several quantum key distribution (QKD) protocols~\cite{LoMDI13, Pan_roundrobin}, which are provably secure techniques that allow two spatially remote users (Alice and Bob) to share a random secret string in the presence of an eavesdropper (Eve)~\cite{BarnettBook}. 

A typical two-photon interference-based QKD scheme, such as the measurement device-independent QKD protocol~\cite{LoMDI13}, requires Alice and Bob to transmit quantum states to a third party (Charlie), who interferes the photons at a beamsplitter, records the time-of-arrival using single-photon counting detectors and announces the detection statistics~\cite{TittelMDI2017}. If the two photons are indistinguishable, they always leave the beamsplitter from the same output port (ideal case), resulting in no coincidence events in the output detectors. This effect, which arises from the destructive interference of the photons' probability amplitudes, is known as the two-photon Hong-Ou-Mandel (HOM) interference~\cite{1987_PRL_HOM}. Observing a coincidence in the two output detectors indicate that the quantum states are distinguishable, and therefore can be used to bound the disturbance caused by an eavesdropper in the quantum channel. 

Because coincidence counts can be used to determine the degree of distinguishability, a two-photon interference-based measurement scheme can also be used in a prepare-and-measure scenario in which the measurement is performed by the receiver (Bob). Recently, interference schemes have been used in qubit-based (dimension $d = 2$) QKD protocols, such as the round-robin scheme~\cite{Pan_roundrobin}. Generally, a two-photon interference-based scheme in a qubit-based protocol is more complicated than a direct measurement scheme where an incoming photon is measured using a receiver comprising of linear optics and single-photon detectors. A two-photon interference scheme requires a second source of quantum states, single-photon detectors, coincidence counters, etc., making it more complicated than a direct measurement scheme. Unless the QKD scheme provides some additional advantages, such as more stringent security, higher secret key rate, or longer distance communication, a direct measurement scheme is generally preferred. 

Yet, there are several QKD protocols, such as the round-robin~\cite{sasaki2014practical}, high-dimensional ~\cite{Brougham13}, and Chau-15~\cite{Wang15}, where a two-photon interference measurement scheme is potentially easier to implement than the complicated interferometric measurement scheme that is most commonly used. Furthermore, a two-photon measurement scheme provides a means to scale the encoding dimension beyond small $d$ at the cost of  essentially no additional changes to the experimental setup. Specifically, for time-phase encoding, it is possible to change the dimension of the time-bin encoding states using software changes without changing anything in the hardware platform.

Here, we consider a two-photon interference-based measurement, which we refer to as quantum-controlled measurement scheme, as a means to detect $d$-dimensional phase states in a time-bin high-dimensional QKD protocol~\cite{Brougham13}. We call this a quantum-controlled scheme because the measurement process is mediated by the controlled (ancilla) photonic state from Bob's local oscillator, analogous to a control bit in a quantum register circuit~\cite{Quantum_multimeter}. In a time-phase QKD protocol, the time-bin states, denoted as $|t_m\rangle,~m \in \{0,.., d-1\}$ are used to encode information and the corresponding phase states $|f_n\rangle = 1/\sqrt{d} \sum\limits_{m= 0}^{d-1} e^{2\pi i n m/d}|t_m\rangle,~n = 0, ..., d-1$ are used to monitor the presence of an eavesdropper. Each quantum state~\textemdash~time or phase~\textemdash~occupies $d$ temporal bins each of width $\tau$ (typically $\sim100-1,000$~ps) and encodes $\log_2d$ bits of information per photon. 

When the dimension of the system is small, the rate of state preparation $1/(\tau d)$ is much higher than the maximum rate at which most single-photon counting detectors can operate, which is mainly limited by the long detector recovery time $\tau_D$ ($\sim10-100$ ns)~\cite{Marsili13}. During this window, a single-photon detector can only detect one incoming photon, thereby limiting the rate of photon detection, an effect known as detector saturation. High-dimensional time-phase encoding ($d > 2$) alleviates this problem by encoding more than one bit of information per photon. By tuning the dimension of the encoding states and matching the expected rate of photon detection $\sim \eta/(\tau d)$ to 1/$\tau_D$, where $\eta$ is the global system loss, it is possible to maximize the bits of information per received photon and hence the secret key rate. Using this technique, many research groups~\cite{MIT16,Taimur2017} have achieved high-secret key rates, demonstrating the feasibility of high-dimensional QKD, thereby overcoming the detector saturation problem.  

One challenging aspect of this protocol is that generating the phase states requires substantial experimental resources. When $d$ is small, the phase states can be generated using a few arbitrary pattern generators, digital-to-analog converters and phase modulators. As $d$ increases beyond small $d$'s, generating the phase states become more expensive and challenging. To solve this problem, we recently studied the feasibility of using only a subset of the $d$ phase states, and demonstrated that the protocol can be secured with just one phase state, although at the cost of lower error tolerance~\cite{Taimur2018}. An implication of this result is that the dimension of the encoding states can be changed using only software. As an example, consider a time-phase QKD system where Eve's presence is monitored by transmitting and measuring the state $|f_0\rangle$, which does not require phase modulation on individual time bins (see above). Then, the the dimension of the QKD system can be changed by redefining the the time and phase states at the software level prior to the communication, without any changes to the physical transmitter set-up.


Another challenging aspect of this protocol is that measuring the phase states requires complicated measurement schemes, such as a combination of electro-optic modulators and fiber Bragg gratings~\cite{Lukens18}, or a tree of time-delay interferometers (DI)~\cite{Taimur2017}. Our past work has primarily focused on using a tree of time-delay interferometers to measure the phase states~\cite{Taimur2017}. Some drawbacks of the interferometric scheme are that the number of time-delay interferometers required to detect a $d$-dimensional phase state scales as $2d - 1$, and the efficiency of the state detection decreases as $1/d$~\cite{Taimur2017}. As a result, the protocol is difficult to scale beyond small $d$. 

To establish a truly arbitrary dimensional QKD, here we propose and implement a quantum-controlled measurement scheme based on two-photon interference that can be used to detect a phase state of any dimension.

The rest of the paper is organized as follows: In Sec.~\ref{equivalence}, we discuss qualitatively the equivalence of a two-photon interference with the interferometers-based measurement scheme. In Sec.~\ref{security-discussion}, we discuss the security analysis of the protocol that we use to analyze the data in a proof-of-principle demonstration of the protocol (Sec.~\ref{Proof-of-principle}). Finally, we conclude the paper in Sec.~\ref{Conclusion} with a perspective on future improvements.  


\section{Quantum Controlled Measurement Scheme}\label{equivalence}
\begin{figure}
\begin{center}
    \includegraphics[width=\linewidth]{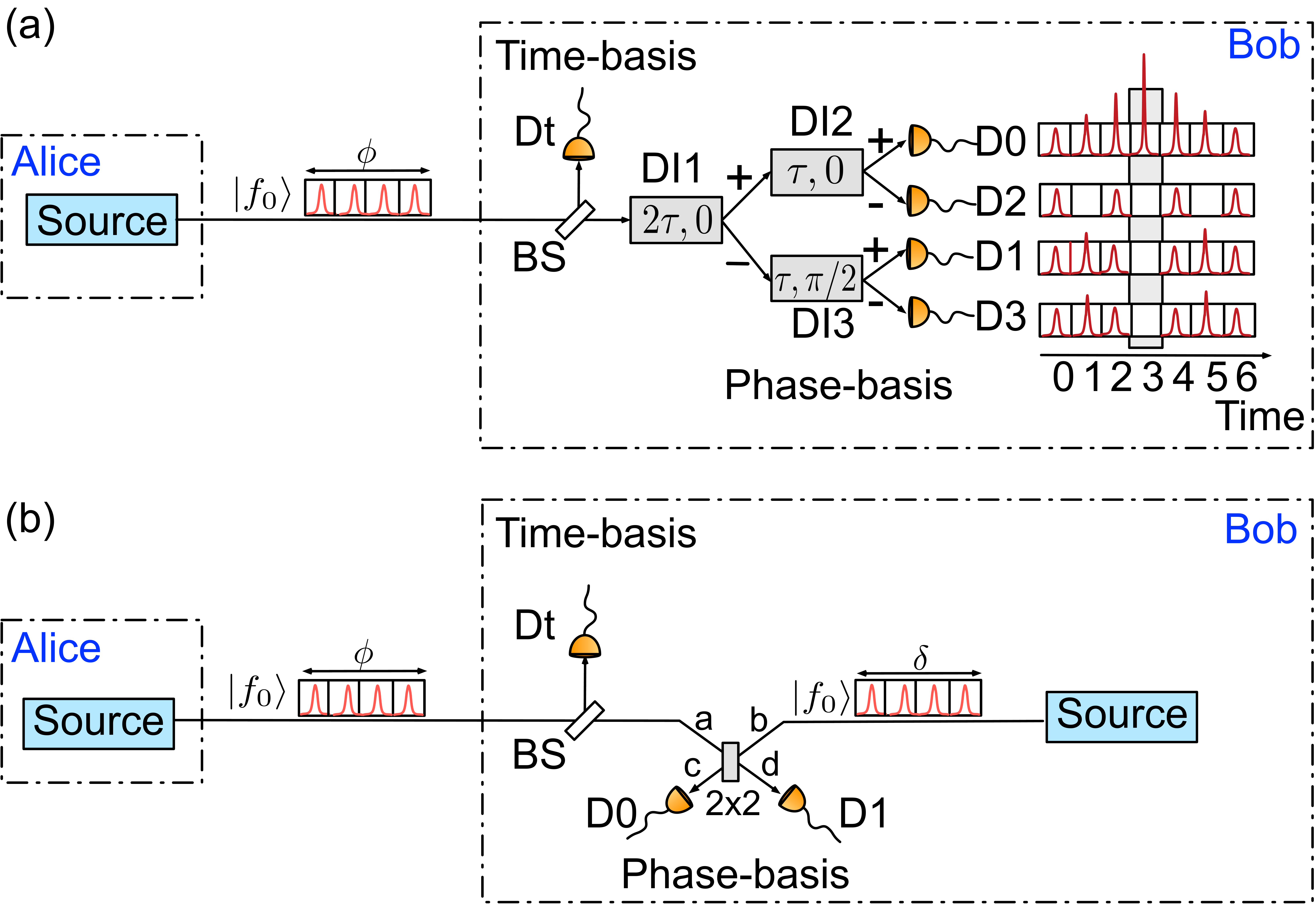}
		\caption{\textbf{Illustration of different phase-state measurement schemes.} A phase measurement scheme consisting of (a) tree of time-delay interferometers and (b) free-running indistinguishable sources.}
		\label{LOScheme}
	\end{center}
\end{figure}

In $d$-dimensional time-phase QKD, the time and phase states are prepared with biased probabilities $p_\textsf{T}$ and $p_\textsf{F}:= 1 - p_\textsf{T}$, respectively. When the quantum states arrive in Bob's receiver, a beamsplitter is used to randomly direct the incoming states for temporal or phase basis measurement. The time-basis states are detected using a single-photon detector connected to a high-resolution time-to-digital converter, and the phase states are measured either using a tree of time delay interferometers or the quantum controlled scheme as we described below and illustrated in Fig.~\ref{LOScheme}.

Consider the $d$-dimensional phase state $|f_0\rangle = 1/\sqrt{d} \sum_{i = 0}^{d-1} |t_i\rangle$, which is used as a monitoring basis state in an asymmetric time-phase QKD protocol. Experimentally, the state $|f_0\rangle$ can be generated by modulating a continuous-wave laser into a wavepacket consisting of narrow-width peaks in $d$ contiguous temporal bins. The overall phase of the wavepacket, denoted as $\phi$, is randomized between each transmission attempt by Alice, but the local phase between the peaks remains coherent with a phase difference taken to be zero. To measure a $d = 4$ phase state, the tree-like arrangement consists of three time-delay interferometers (DIs) with the optical path-length differences and the phases of the interferometers set so that there is a one-to-one mapping between the input phase state $|f_n\rangle$ and the detector D$n$ in which the event is registered~\cite{Taimur2016}, as shown in Fig.~\ref{LOScheme}(a). 

A qualitative analysis of the interference pattern observed in the output detectors reveals that the effective function of the interferometric tree is to delay each successive peak, and interfere them all in the same time-bin, resulting in a constructive interference in time bin 3 at the output detector D0 (right panel of Fig.~\ref{LOScheme}(a)). When the incoming state is the ideal $|f_0\rangle$ state, no event is recorded in the central time bins of detectors D1-D3 due to the destructive interference of the wavepacket peaks. Any disturbance of the incoming state results in incomplete destructive interference, leading to events appearing in time bin 3 in detectors D1-D3, and hence can be used to identify eavesdropping. The probability amplitudes observed in all other time bins in all the detectors result from the interference of a subset of the wavepacket peaks and do not provide complete information about the incoming quantum state. See Ref.~\cite{Taimur2016} for a detailed description of how the interferometric setup detects the phase states. 


An equivalent scheme to detect the phase state is to interfere the incoming state from Alice with a locally generated state in the receiver. Bob's source can have an arbitrary phase $\delta$ with respect to $\phi$, as long as his source generates quantum states that matches Alice's in the spatial, spectral, polarization and temporal domains. When the phase of Bob's quantum state is arbitrary with respect to Alice's, the resulting interference pattern shows the HOM effect. Thus, disruption of the HOM two-photon interference can be also be used to detect any perturbation of the incoming state from Alice as illustrated in Fig.~\ref{LOScheme}(b). 

To see how HOM interference can be used to determine perturbation of Alice's states in the quantum channel, we consider the simple case where the incoming state from Alice is a pure single-photon state, has local phase perturbations and is input through the beamsplitter port $a$. The incoming state can be represented as
\begin{align}
|f_0\rangle_A = \frac{1}{\sqrt{d}} \sum_{i = 0}^{d-1} e^{i \phi} e^{i \lambda_i} |t_i\rangle = \frac{1}{\sqrt{d}} \sum_{i = 0}^{d-1} e^{i \phi} e^{i \lambda_i} a_i^{\dagger}|\text{vac}\rangle,
\end{align}
where $e^{i \lambda_i}$ is a complex number of unit magnitude, the operator $a_i^{\dagger}$ denotes the field creation operator in temporal mode $i$, and $|\text{vac}\rangle$ represents the vacuum. We assume that Bob's locally generated state is ideal, is coupled into the beamsplitter through the input port $b$, and is described by
\begin{align}
|f_0\rangle_B = \frac{1}{\sqrt{d}} \sum_{i = 0}^{d-1} |t_i\rangle = \sum_{i = 0}^{d-1} b_i^{\dagger}|\text{vac}\rangle. 
\end{align}
A beamsplitter transforms the creation operator $a_i^{\dagger}$ ($b_i^{\dagger}$) as $1/\sqrt{2}(c_i^{\dagger} + d_i^{\dagger})$ [$1/\sqrt{2}(c_i^{\dagger} - d_i^{\dagger}$)]. Therefore, the overall transformation of the input states $|f_0\rangle_A |f_0\rangle_B$ through the beamsplitter can be written as

\begin{align} \label{BS_Output} 
    |f_0\rangle_A |f_0\rangle_B &\rightarrow \frac{1}{2d} \sum_{i, j = 0}^{d-1} e^{i\phi} e^{i\lambda_i} (c_i^\dagger + d_i^\dagger)(c_j^\dagger - d_j^\dagger)|\text{vac}\rangle\\
   = \frac{1}{2d} \biggl\{ &\sum_{i = j = 0}^{d-1} e^{i\lambda_i} (c_i^{\dagger} c_i^{\dagger} -  d_i^{\dagger} d_i^{\dagger})  +  \nonumber \\ 
    &\sum_{i \neq j = 0}^{d-1} [(e^{i\lambda_i} + e^{i\lambda_j}) (c_i^{\dagger} c_j^{\dagger} -  d_i^{\dagger} d_j^{\dagger}) + \nonumber \\ 
    &(e^{i\lambda_i} - e^{i\lambda_j}) (c_j^{\dagger} d_i^{\dagger} -  d_j^{\dagger} c_i^{\dagger})] \biggr\} |\text{vac}\rangle.
\end{align}

An important property of the output state in Eq.~\ref{BS_Output} is that the probability of observing a coincidence in the two output detectors ($|e^{i\lambda_i} - e^{i\lambda_j}|^2$) goes to zero when the incoming state is ideal ($\lambda_i = \lambda_j = 1$). Any phase perturbation that results due to an eavesdropper trying to estimate the incoming quantum states manifests itself in the form of coincidence with probability $|e^{i\lambda_i} - e^{i\lambda_j}|^2$, which can be detected in the experiment as quantum bit errors in the phase basis. This is similar to the interferometric scheme where the perturbations result in events in the central time bins in output detectors D1-D3, as opposed to no expected events due to destructive interference in the absence of a phase perturbation. 

This simple example gives an intuitive explanation of how the quantum-controlled scheme can be used to monitor the presence of an eavesdropper. Below we extend this to the case where Alice transmits phase-randomized weak coherent states and Eve can attack the quantum states using collective attack strategies.


\section{Security Analysis}\label{security-discussion}
The security of this protocol is based on the semi-definite programming (SDP)-based proof we presented in Ref.~\cite{Taimur2018}. Here, for completeness, we briefly summarize the security analysis, starting with the case where Alice sends pure single-photon states and later extend it for the more practical case of weak coherent states with three-level decoy wavepackets. Using the decoy states, we place bounds on the single-photon detection rates in the time and phase basis. We also show a new approach to bound the single-photon error rates in the phase basis, inspired from the two-photon interference technique derived in Ref.~\cite{AndresOL}. The novelty of this technique is that it allows us to bound the two-photon error rate in the phase basis $|\lambda_i - \lambda_j|^2$ using only two decoy states, and single-click statistics from the two single-photon detectors. We have modified the approach in Ref.~\cite{AndresOL} to account for the arbitrary photon number distribution that Eve can inject into the receiver. Our decoy-state approach can also be used in other two-photon-based interference schemes such as measurement-device-independent (MDI) QKD~\cite{TittelMDI2017}.

For the security analyses presented below, we assume that Alice transmits an infinitely long bit string (block size $N\rightarrow{\infty}$), ignoring the finite-key effects for now, which are known to be significant if the block size is shorter than $\sim 10^6$ bits~\cite{CharlesDecoy}. Later, we provide some insights on how to consider the finite-key length, which we will address in a future work. 

\subsection{Semi-definite Programming}
We assume that Alice transmits single-photon states of arbitrary dimension in both the time and phase basis. In the time basis, Alice transmits $d$ temporal states $\{t_1, ..., t_{d-1}\}$, and in the phase basis she only transmits the state $\{f_0\}$. A common technique to analyze the security of a prepare-and-measure protocol is to represent the states in an equivalent entanglement-based picture, assume Eve interacts with each quantum states collectively, and then promote the analysis to general attacks using known techniques such as de Finetti theorem~\cite{Renner05,Renner07,Renner09,RennerdeFinetti09}. 

In the equivalent entanglement-based picture, Alice and Bob share an entangled state of the form $|\phi_{\textsf{X}}\rangle = 1/\sqrt{d} \sum_{i = 0}^{d -1} |x_i\rangle_A |x_i\rangle_B$ where $x~\in~\{t, f\}$ and $\textsf{X}~\in~\{\textsf{T}, \textsf{F}\}$. The density matrix of the entangled state can be represented as $\rho_{A,B} = |\phi_{\textsf{X}} \rangle \langle \phi_{\textsf{X}}|$. Eve's collective interaction with the entangled state transforms the state into $|\Psi_{\textsf{X}} \rangle_{A,B,E} = \sum_i \sqrt{\gamma_i} |\phi_{\textsf{X}} \rangle |\gamma_i\rangle_E$, where we assume her interaction with the state is independent and identically distributed (i.i.d). Bob's measurement operators for each basis state is defined as $\Pi_{\textsf{X}_i} = |x_i\rangle \langle x_i|$. For the detection events that he classifies as error bits, the operators are defined as $E_{\textsf{X}}$ corresponding to quantum bit error rates $e_{\textsf{X}}$~\cite{Taimur2018}.

The main goal of the security proof is to bound the hypothetical error rate (the so-called phase error rate $e_{\textsf{F}}^{\mathsf{U}}$) when the entangled state is in the phase basis, but both Alice and Bob perform measurements using the time-basis operator. This hypothetical error rate is different from the quantum bit error rates in the time $e_{\textsf{T}}$ and the phase basis $e_{\textsf{F}}$ as described above in that the former is quantified as a bound and the latter are determined from the experiment. Throughout the rest of the paper, we distinguish these by referring to them as the \textit{phase error rate} and the quantum bit error rates in the time and phase basis.

The SDP-based security analysis allows us to quantify $e_\textsf{F}^{\mathsf{U}}$ under the assumption that the quantum bit error rates for each basis are known \textit{a priori}. Formally, the problem becomes
\begin{align}
    \texttt{maximize}~~\text{Tr}~[E_\textsf{F} \rho_{AB}] &= e_\textsf{F}^{\mathsf{U}} \\
    \texttt{s.t.} \text{Tr}~[\rho_{AB}] &= 1,~~\rho_{AB} \geq 1, \\
    \text{Tr}~[E_\textsf{T} \rho_{AB}] &= e_\textsf{T}, \\
    \text{Tr}~[\Pi_{x_i} \otimes \Pi_{y_j}] &= p_{i,j} \label{upper-bound}\\
    \forall \{x,y\} \in \{\mathsf{T}, \mathsf{F}\}~~~~&\&~~~~i,j = 0, ..., d-1. 
\end{align}
Here, we make two assumptions regarding Bob's phase state measurement device. First, we assume that Bob's device is ideal which can be adapted to the practical, non-ideal case by changing the values of $p_{i,j}$. Specifically, by preparing and measuring states in both the time- and phase-basis, the values of $p_{i,j}$'s can be set to the probabilities estimated from the experiment. Second, we assume that Bob's phase measurement scheme is well calibrated. In the context of this protocol, this means that the Bob generates phase states $|f_0\rangle$ with well-defined phase values on each wavepacket peak. This is a valid assumption in a prepare-and-measure scheme because Eve does not have access to Bob's measurement devices.

An interesting feature of our security analysis is that it is highly flexible and allows us to change the dimension, number of monitoring states, and probability with which Alice wants to transmit these states easily through a simple program. We use the CVX~\cite{cvx} library in Matlab to optimize the value of $e^{\mathsf{U}}_\textsf{F}$. For additional information regarding this approach, we refer readers to Ref.~\cite{Taimur2018}. We also make this code available for the community through Ref.~\cite{code}.

Our security analysis is not only valid for this protocol, but also for any other two-basis QKD protocol with a direct measurement scheme, \textit{e.g.}, high-dimensional spatial-modes-based scheme~\cite{Mirhosseini_2015}. To use our program for any other protocol, one has to estimate the values of $e_\mathsf{T}$ (assuming symmetric error rates in both the basis, \textit{i.e.}, $e_\mathsf{T}$ = $e_\mathsf{F}$), and $p_{i,j}$'s from experimental calibration. Using these as \textit{a~priori} known statistics, one can then solve for $e^\mathsf{U}_\mathsf{F}$ using our Matlab program. Additionally, if Alice and Bob only use a fraction of the states as monitoring states as we do here by transmitting only $|f_0\rangle$, it is possible to calculate $e^\mathsf{U}_\mathsf{F}$ by appropriately adjusting the combinations in Eq.~\ref{upper-bound}.

\subsection{Decoy State Formalism}
The SDP-based security analysis discussed above is formulated under the assumption that Alice and Bob transmit ideal single-photon states. In practice, most QKD systems are implemented using attenuated coherent laser sources that generate photons based on a Poisson distribution. Due to the probabilistic nature of the source, the phase randomized weak coherent states (PRWCS) generated from a coherent laser includes, in addition to the single-photon states, vacuum and multi-photon states. A commonly used technique to overcome this problem is to generate PRWCS with different mean photon numbers, and use the detection rates of each mean photon number to quantify the fraction of the states that contains exactly 1 photon, also known as the decoy state method~\cite{WangDecoy_1,Lodecoy05}.

Due to the nature of this protocol, implementing decoy-state methods for our $d$-dimensional protocol is slightly complicated. The main challenge is that the time-basis states are measured using a direct measurement scheme, while the phase-basis states are measured using a two-photon interference scheme. For the time-basis, we can adapt the existing three mean photon number decoy bounds that are commonly used in most prepare-and-measure QKD schemes. For the phase-basis, we identify a technique to bound the single-photon error rates directly from the click statistics measured in the experiment~\cite{AndresOL}.

To implement the decoy method, we assume that Alice transmits quantum states of three different mean photon numbers $\mu_1, \mu_2$ and $\mu_3$ where $\mu_2 + \mu_3 < \mu_1$ with probabilities $p_{ \mu_1}, p_{\mu_2}$ and $p_{\mu_3}$, respectively. For the phase measurement scheme, Bob also generates the phase basis state $|f_0\rangle$ with mean photon numbers $\mu_1, \mu_2$ and $\mu_3$. Under these assumptions, the secret key rate of the system can be written as~\cite{Valerio09}
\begin{align}\label{eq:SKR}
    r := R_{\textsf{T},1}[\log_2 d - H(e^{\mathsf{U}}_\textsf{F})] - \Delta_\textsf{EC},
\end{align}
where $r$ is the secret key rate, $R_{\textsf{T},1}$ is the single-photon gain in the time-basis (see below), $H(x):= -x\log_2 (x/d) - (1-x) \log_2(1-x)$ and $\Delta_\textsf{EC}:= R_\textsf{T} H(e_{\textsf{T}})$ represents the number of bits used in the error correction with $R_\textsf{T}$ being the detection rate in the time-basis.

In this protocol, the phase error rate $e^\mathsf{U}_\textsf{F}$ depends on the quantum bit error rate in the phase basis $e_\textsf{F}$, which is defined as the fraction of the events that are recorded as coincidence events in the detectors D0 and D1 when single-photon states are input from the ports $a$ and $b$ of the beamsplitter. Estimating $e_\textsf{F}$ is somewhat complicated because here both Alice and Bob transmit PRWCS, which means that we have to bound them based on the detection statistics to estimate the conditional probabilities of receiving single-photon state from each Alice and Bob. There are three primary steps to bounding $e^\mathsf{U}_\textsf{F}$, which are also provided in the supplementary information in detail. Here, we briefly summarize these steps for completeness.

First, we upper-bound the conditional probability $Y^\mathsf{U}_{11}$ that Alice and Bob each transmit a single-photon state and both detectors D0 and D1 record a detection, resulting in a coincidence event. These can be estimated from the coincidence probabilities $C^{\mu_i,\mu_j}$ where only four combinations of $\mu_i$ and $\mu_j$ are necessary (see Supplementary Information Sec. 1). We calculate the upper-bound of the conditional coincidence probability as

\begin{align}
Y^\mathsf{U}_{11} \leq \frac{1}{\mu_i \mu_j}\left[ C^{\mu_i,\mu_j}e^{\mu_i + \mu_j}-(C^{\mu_i,0}e^{\mu_i}+C^{0,\mu_j}e^{\mu_j})+Y_{00} \right],
\label{eq_Y11}
\end{align}
where $Y_{00}$ represents the conditional probability of observing a coincidence when both Alice and Bob transmit vacuum states.

Second, we estimate the lower-bound conditional probability $Y^\mathsf{L}_{\mathsf{F_A},1}$ ($Y^\mathsf{L}_{\mathsf{F_B},1}$) of Alice (Bob) transmitting a state with a single photon and either of the detectors D0 or D1 recording a detection event. These can be estimated using the events where Alice (Bob) transmits state $|f_0\rangle$ and Bob (Alice) transmits a vacuum (see Supplementary Information Sec. 1). These detection rates allow us to lower bound the fraction of input states to the beamsplitters that contains exactly one photon at each input port $a$ and $b$. Using these lower bounds, we can calculate the quantum bit error in the phase basis as
\begin{align}
    e_\textsf{F} \leq \frac{Y^\mathsf{U}_{11}}{Y^\mathsf{L}_{\mathsf{F_A},1} Y^\mathsf{L}_{\mathsf{F_B},1}}.
\end{align}
Finally, we use the SDP program and the value of $e_\textsf{F}$ to calculate the $e^{\mathsf{U}}_\textsf{F}$. Based on the results from Ref.~\cite{Taimur2018}, we note that $e_\textsf{F} = e^{\mathsf{U}}_\textsf{F}$ if Alice and Bob transmit all $d$ or $d-1$ phase states as is required for a complete protocol. Since only the state $|f_0\rangle$ is transmitted as the monitoring basis state in this protocol, we use $e_\textsf{F}$ as the \textit{a priori} error rate in the SDP program to estimate $e^{\mathsf{U}}_\textsf{F}$.

In the time-basis, the single-photon gain, defined as the joint probability that Alice transmits a time-basis state and Bob receives a detection click, is bounded by
\begin{align}\label{single-photon-gain}
R_{\textsf{T},1} = [p_{\mu_1} (\mu_1 e^{\mu_1})+p_{\mu_2} (\mu_2 e^{\mu_2}) + p_{\mu_3} (\mu_3 e^{\mu_3})]Y_{\textsf{T},1}.
\end{align}
In Eq.~\ref{single-photon-gain}, $Y_{\textsf{T},1}$ represents the conditional probability that Bob's detector records an event given Alice transmits a single-photon state~\cite{PracticalDecoy05}
\begin{align}\label{eq:single_photon_yield}
&Y^\mathsf{L}_{\mathsf{T},1}  \geq \frac{\mu_1}{\mu_1 \mu_2 - \mu_1 \mu_3 - \mu_2^2 + \mu_3^2} \times \nonumber \\
&\biggl[ R_{\mathsf{T},\mu_2} e^{\mu_2} - R_{\mathsf{T},\mu_3} e^{\mu_3} - \frac{\mu_2^2 - \mu_3^2}{\mu_1^2}(R_{\mathsf{T},\mu_1} e^{\mu_1} - Y_{\mathsf{T},0})\biggr],
\end{align}
where $R_{\mathsf{T},\mu_i}~i\in\{1,2, 3\}$ represents the joint probability of Alice transmitting a state with mean photon number $\mu_i$ and Bob receiving a detection event,  and the term $Y_{\mathsf{T},0}$ is the zero-photon yield defined as
\begin{align}
Y_{\mathsf{T},0}:= Y^\mathsf{L}_{\mathsf{T},0} \geq \frac{\mu_2 R_{\mathsf{T},\mu_3}e^{\mu_3} -
\mu_3 R_{\mathsf{T},\mu_2}e^{\mu_2}}{\mu_2 - \mu_3}.
\end{align}

With all terms defined, we can now derive the phase-error-rate bound and simulate the secret key rate as defined in Eq.~\ref{eq:SKR}. To derive the upper bound $e^{\mathsf{U}}_\textsf{F}$, we use $e_\textsf{F}$ as the quantum bit error rate in the phase basis and use the SDP approach to calculate $e^{\mathsf{U}}_\textsf{F}$. This bound is generally worse in the case where Alice transmits just one state in the phase-basis than the case where Alice and Bob transmit all $d$ or $d-1$ phase states~\cite{Taimur2018}. Despite the worse bound, transmitting just one phase state is much simpler to implement experimentally and easier to scale. If the HOM interference visibility between Alice and Bob's source is high, the difference between sending one or $d$ states is small, which can lead to high secret key rate in an experiment as we showed below. 

\section{Experimental Demonstration}\label{Proof-of-principle}
\begin{figure*}[htb]
\begin{center}
    \includegraphics[width=0.7\linewidth]{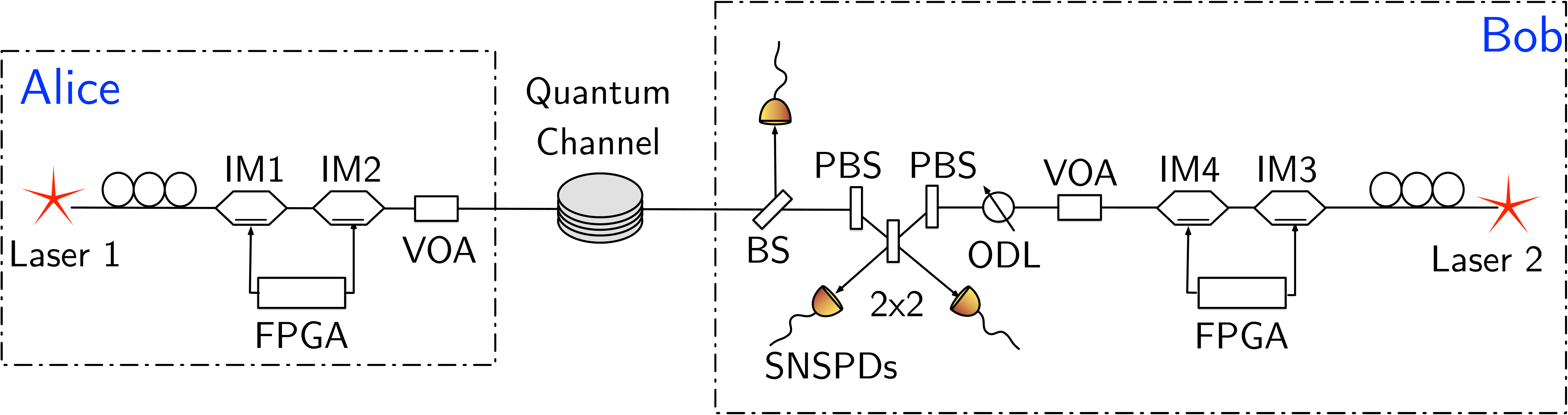}
		\caption{\textbf{Experimental setup}. A field programmable gate array (FPGA, Altera Stratix V) with 7 independent channels is used to drive Alice and Bob's intensity modulators (EOSpace) to generate the time and phase basis states. After the quantum channel, photons incoming to Bob's receiver are passed through a fiber beamsplitter that makes a passive basis choice.}
		\label{Experimental_Setup}
	\end{center}
\end{figure*}

Our proof-of-principle experimental setup is shown in Fig.~\ref{Experimental_Setup}. Alice's transmitter consists of two electro-optic intensity modulators driven with a field programmable gate array (FPGA) to create the time and phase states at a rate of $2500/d$ MHz. The first intensity modulator (IM1) is used to generate the temporal states $\{t_0, t_1, ..., t_{d-1}\}$ and the phase state $\{f_0\}$ with an optical pulse width of $\sim$80~ps in a single temporal bin of width 400~ps. The second intensity modulator (IM2) is used to normalize the amplitude level of the phase state $\{f_0\}$ and to create the signal, decoy and vacuum states with mean photon numbers $\mu_1$, $\mu_2$ and $\mu_3$, respectively. A variable optical attenuator (VOA) is used to reduce the mean photon number, red and a second one is used to simulate the quantum channel loss. Although the extension of this protocol to a real fiber is known to be straightforward, there are certain challenges that need to be considered. In a fiber implementation, the primary challenge will be the drift in polarization and phase, neither of which will have significant impact on the measurement of the time basis states. For the phase-basis measurement, the phase drift does not have any impact because the information is encoded as the relative phase difference between the wavepacket peaks on a time scale of $\tau d$, which is much shorter than the typical time scale of phase changes in a fiber from environmental factors. The drift in polarization can  only impact the indistinguishability of the photon which is  negated  by  placing  a  polarizing  beamsplitter (PBS) at the input of Bob's receiver (see below). In this case, the impact of the polarization drift will be equivalent to a drift of channel loss in both the time and phase bases. 

A section of Bob's receiver has an identical layout as Alice's setup in which the intensity modulators (IM3 and IM4) are used to generate the phase state $|f_0\rangle$ with the same mean photon numbers as Alice. Bob's quantum states are then attenuated to match the detection rate of the incoming quantum states from Alice. The time-basis measurement is performed using a superconducting nanowire single-photon detector (SNSPD, Quantum Opus) with a nominal detection efficiency of 80\%, timing jitter of 50 ps, dark count rates of $<$ 100 cps and a deadtime of $\sim$~30 ns. For the phase-basis quantum-controlled measurement, both Alice and Bob's states are passed through two fiber-based PBSs before interfering at a 50/50 polarization maintaining fiber beamsplitter (2$\times$2). Although we place the first PBS after the BS, to ensure that Eve cannot take any advantage of the polarization mismatch between the two bases, it could also be placed before the BS without affecting the system operation. The two output ports of the beamsplitter are coupled into two nominally identical SNSPDs. The detector signals are time-tagged using a time-to-digital converter (Agilent Acqiris U1051A) and streamed to a computer for further post-processing. 

The indistinguishability of Alice and Bob's quantum states in the spectral domain is ensured by mixing a fraction of the power from each output beam of their lasers and generating a beatnote frequency, which is detected using a high-speed photoreceiver (Miteq DR-125G-A, not shown in Fig.~\ref{Experimental_Setup}). Assuming the laser center frequencies are within $\Delta \nu$ of each other, the phase between the two lasers increases by $2\pi \Delta \nu t$ as a function of time $t$. Alice's laser (Wavelength Reference Clarity NLL-1550-HP) is locked to a hydrogen cyanide molecular absorption line and Bob's laser (Agilent HP81682A) is tunable within a resolution of 0.1 pm. This allows us to tune the beatnote frequency between the two lasers well below 10 MHz, which is equal to a phase shift of $0.05$ rad for $d = 2$ states. Throughout the experimental runs, the beatnote is monitored periodically and tuned as required. In this work, we do not actively phase randomize Alice and Bob's quantum states; instead, we take advantage of the slight mismatch between the two lasers' center frequencies to randomize the phase. In the future, two high-speed phase modulators, independently driven by two FPGAs can be used to randomly modulate the phase of each state~\cite{Cao2015}.

\begin{figure}
\begin{center}
    \includegraphics[width=\linewidth]{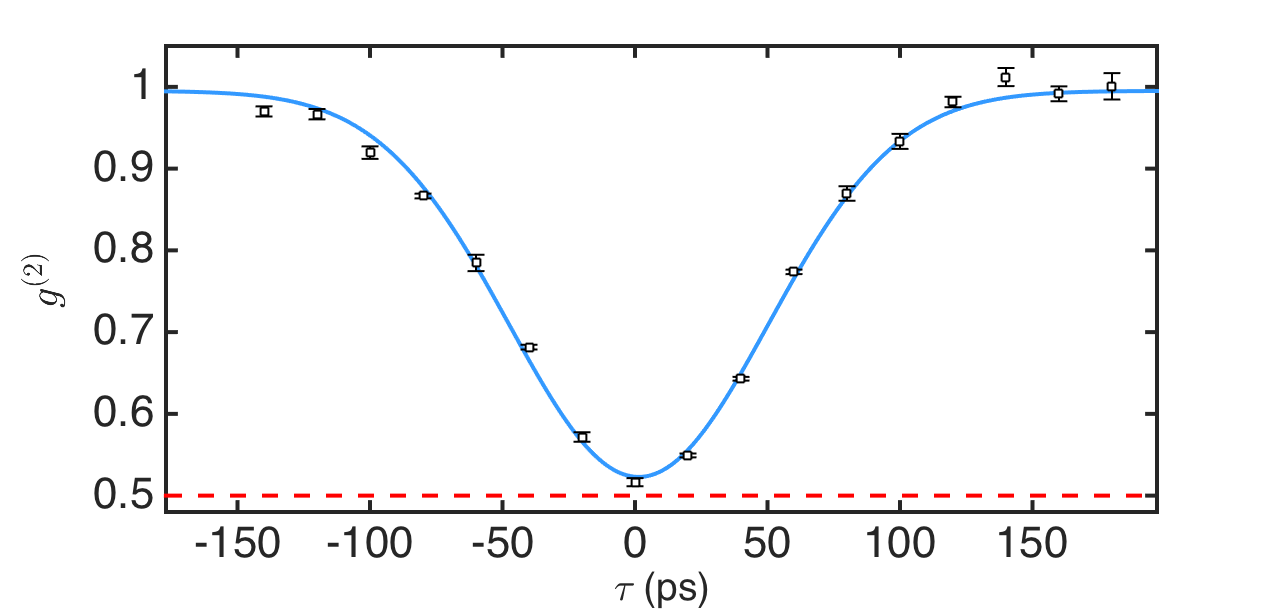}
		\caption{\textbf{Second-order coherence function} The temporal delay is scanned in steps of 20-ps, and the coincidence counts are recorded. The visibility $V \equiv 1 - g^{2}(0)$ is determined to be 0.48 $\pm$ 0.02.}
		\label{Coincidence_Probability}
	\end{center}
\end{figure}

To test the indistinguishability of Alice and Bob's optical wavepackets, we characterize the HOM interference between the two sources based on the second-order coherence function $g^{(2)}(\tau)$ as a function of the relative time-delay $\tau$ between the wavepackets. For the characterization measurement, the intensity modulators are driven with a periodic pattern from the FPGA, which generates single-peaked optical wavepackets at a repetition rate of 7.81 MHz with a mean-photon number of $0.016 \pm 0.001$. The relative temporal delay between the wavepackets is tuned using an optical delay line (ODL, General Photonics VDL-001-35-50-SS-FC/APC) over a temporal range of $340$ ps. The coincidence events are recorded for each temporal delay using a time-to-digital converter. Figure~\ref{Coincidence_Probability} shows a typical HOM interference signature represented as $g^{(2)}(\tau) = C^{\mu, \mu}/(S_1 S_2)$, where $S_1$ and $S_2$ represent the probability of detecting single events in the detectors D0 and D1, respectively. When the wavepackets are completely overlapping, the coincidence counts decreases approximately by a factor of 2, corresponding to a $g^{(2)}(0) = 0.52 \pm 0.02$, which is consistent with the theoretical limit of 0.5 (red dahsed line) within the experimental uncertainties.

A high-visibility HOM interference in the temporal domain can be achieved from nominal mode matching in the polarization, temporal, spectral and spatial domains. During a time-phase QKD session, the HOM interference is measured between phase states $|f_0\rangle$, and can therefore suffer from additional factors, such as drift in the relative phase due spectral mismatch between Alice and Bob's lasers, especially when $d$ is large. As a result of the phase drift due to spectral mismatch, the visibility in the phase domain is always worse than in the time domain (see below).

For the proof-of-principle QKD demonstration, we set the temporal delay so that the coincidence rate is minimum (smallest $g^{(2)}(\tau)$), which ensures high interference visibility defined as $V = 1 - g^{(2)}(0)$. A fixed pattern of length $N = 10^{12}/d$ is transmitted from Alice using a pre-determined random basis choice in the FPGA memory. The total time of communication session is set to 400~s, which is long enough to yield $> 10^8$-bit long secret key. Hence, finite-key effects are negligible in the experiment. The time- and phase-basis states are transmitted with equal probabilities, and each of the three mean photon numbers are transmitted with $1/3$ probability. For each channel loss, we tune the dimension of the encoding states between $d = 2$ to  $16$ in powers of 2. 

\begin{figure}
\begin{center}
    \includegraphics[width=\linewidth]{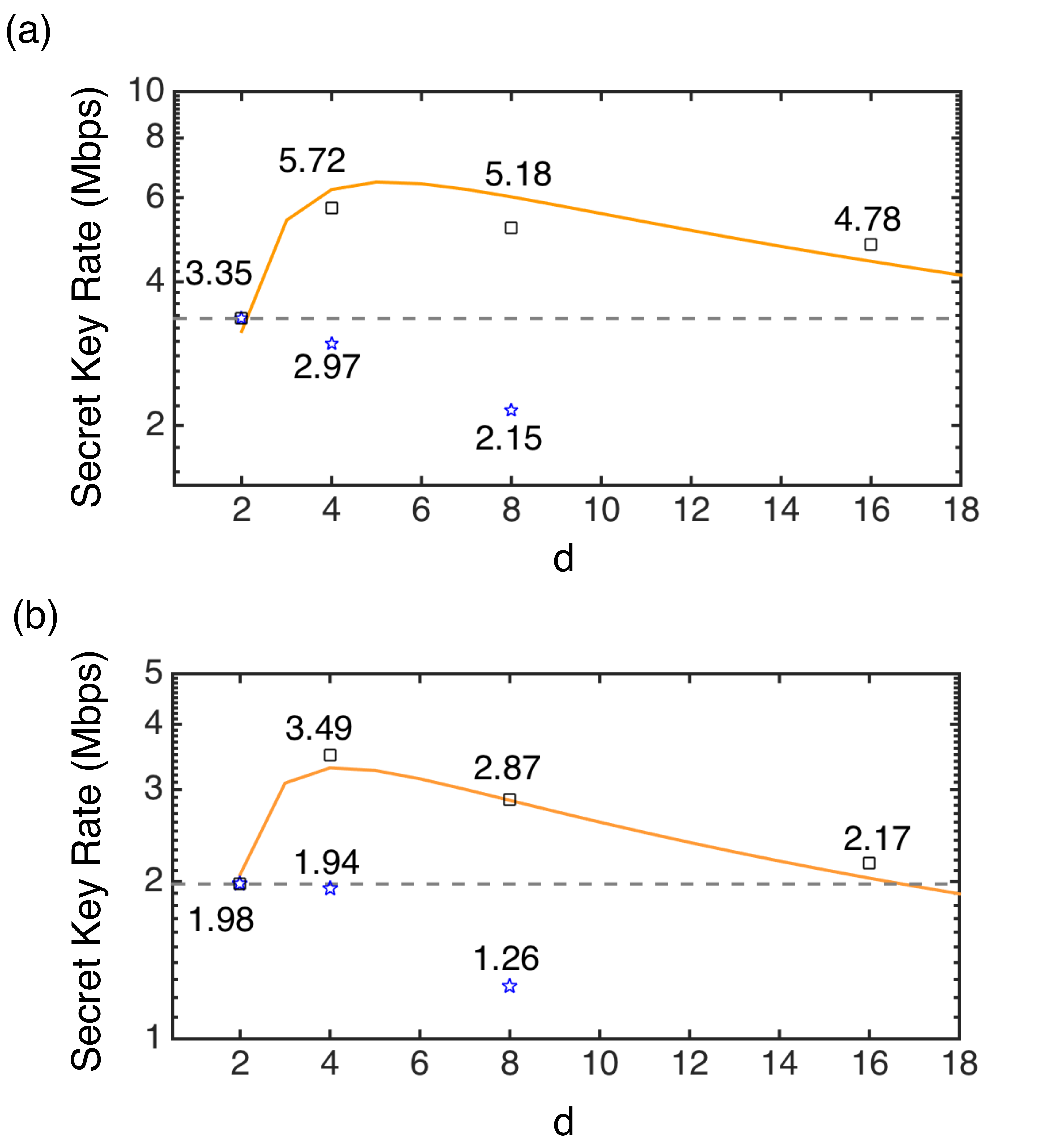}
		\caption{\textbf{Experimental results with $p_\textsf{T} = 0.50$ and $p_\textsf{F} = 0.50$.} Experimentally determined secret key rates (blue pentagrams) plotted as a function of the dimension for a quantum channel loss of (a) $4~$dB  and (b) $8~$dB. The black squares represent the theoretical maximum secret key rate that can be achieved if all $d$ phase states are transmitted with the same error rate ($e_\textsf{F}$) in the phase basis (see Table~\ref{tab:Experimental_Parameters})}. The solid lines show the simulated secret key rates derived using channel model described in Ref.~\cite{Taimur2018, Bing2018}. The dashed lines indicate the secret key rates achieved with $d = 2$ states.
		\label{Experimental_Results}
	\end{center}
\end{figure}

Our main experimental results are shown in Fig.~\ref{Experimental_Results}, where we show the extractable secret key rate as a function of $d$ for two channel losses: 4~dB (Fig.~\ref{Experimental_Results}(a)) and 8~dB (Fig.~\ref{Experimental_Results}(b)). The secret key rate is calculated using Eq.~\ref{eq:SKR}, where the values of $e_\textsf{T},~e_\textsf{F},~e^{\mathsf{U}}_\textsf{F}$, and $\Delta_\textsf{EC}$ are determined experimentally. Table~\ref{tab:Experimental_Parameters} summarizes the experimental parameters, including the mean-photon numbers and the error rates for each photon number $e_{\textsf{T}_{\mu_i}}$ in the time-basis. We have also summarized the definitions of the key symbols in the supplementary material (Sec. 2).

When we assume that $e^{\mathsf{U}}_\textsf{F}$ is equal to $e_\textsf{F}$, we obtain the highest achievable secret key rate for the given error rate (black squares). This is a theoretical limit of the maximum secret key rate that can be achieved if all $d$ or $d-1$ phase states are transmitted~\cite{Taimur2018}. At a channel loss of 4~dB (Fig.~\ref{Experimental_Results}(a)), we observe that this theoretical secret key rate increases with dimension, peaking at $d = 4$, and drops as $d$ is increased beyond $d = 4$. 

An important feature of our SNSPDs is that the detector efficiency changes as a function of detection rate, and the nominal efficiency of $80\%$ is only achieved if the detection rate is $<$1-2 Mcps~\cite{Clinton_RSI_2017}. As the dimension increases beyond $d = 4$, the overall detection rate in the temporal basis drops below 1.72 Mcps, which is approximately below the detector saturation regime of our SNSPDs. At $d = 8$, detector saturation is no longer a dominating factor, and going beyond $d = 8$ to $d = 16$ decreases the maximum achievable secret key rate even more.

When we use $e_{\textsf{F}}$ and the SDP program to calculate the bound $e^{\mathsf{U}}_\textsf{F}$, we find that the secret key rates remains the same (3.35 Mbps) at $d = 2$ but drops to $2.97$ and $2.15$ Mbps for $d = 4$ and $d = 8$, respectively (blue pentagrams in Fig.~\ref{Experimental_Results}(a)). We cannot calculate the bound for $d = 16$ because the dimension of the matrices ($d \otimes d$) goes beyond the maximum size that our version of Matlab can handle on our computing platform.

There are two main reasons why the secret key rate drops rapidly for this case. First, for $d = 2$, the phase error rate for transmitting $d$ or $d-1$ states is same ($e_\textsf{F} = e^{\mathsf{U}}_\textsf{F}$)~\cite{Tamaki2014}, hence there is no penalty for transmitting just one monitoring basis state in $d = 2$. But, as the dimension increases beyond $d = 2$, $e^{\mathsf{U}}_\textsf{F}$ for transmitting fewer than $d-1$ states grows rapidly unless $e_\textsf{F}$ is low. This results in larger overhead to determine the presence of Eve, resulting in a much lower secret key rate. In our experiment, the imperfect HOM visibility leads to a quantum bit error rate in the phase basis ($e_\textsf{F}$) between $\sim3-6\%$, which results in phase error rates between $0.041$ and $0.328$ for both the channel losses, as shown in Table~\ref{tab:Experimental_Parameters}.

\begin{figure}[htb]
    \begin{center}
    \includegraphics[width=\linewidth]{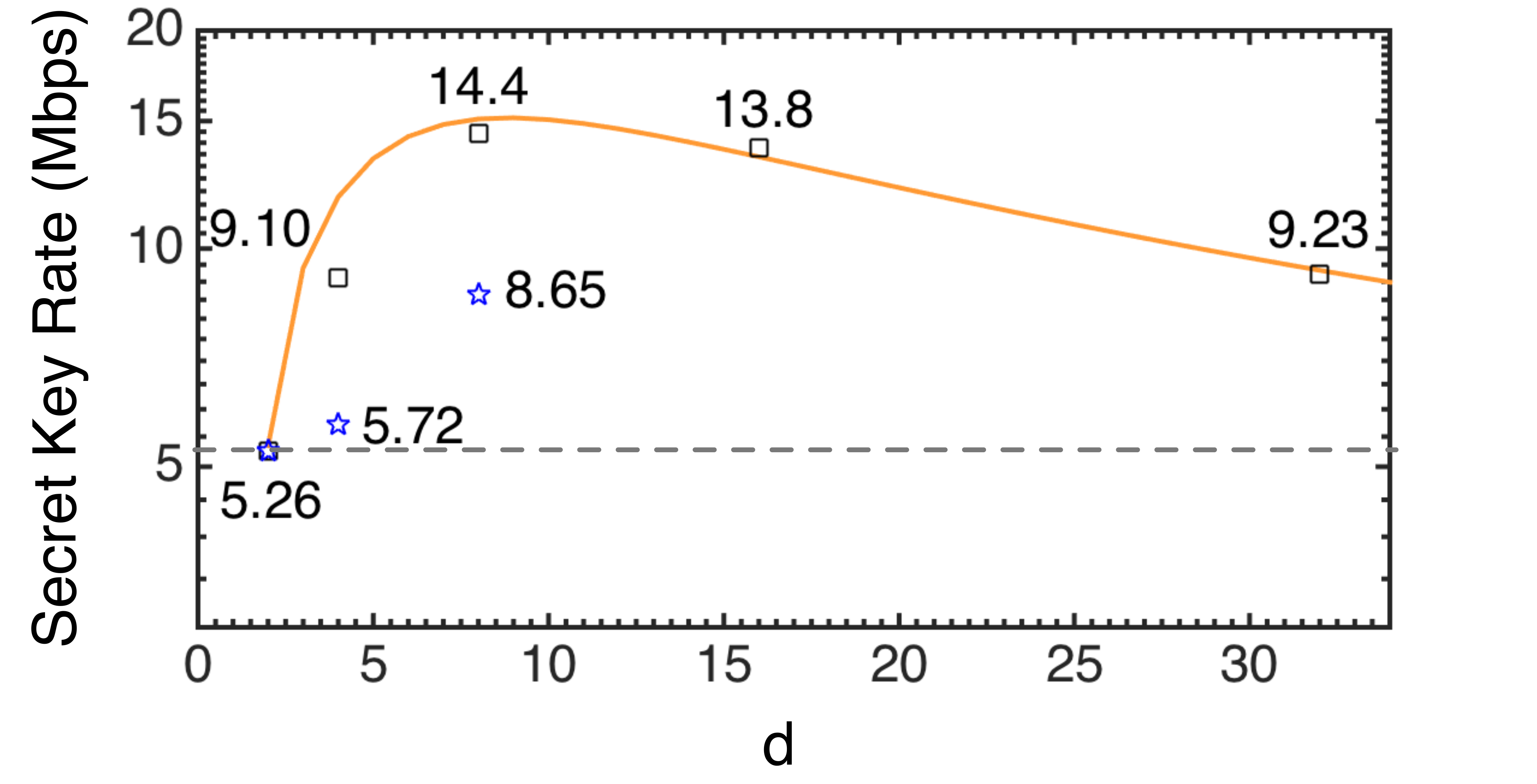}
		\caption{\textbf{Experimental results with $p_\textsf{T} = 0.90$ and $p_\textsf{F} = 0.10$.} Observation of high secret key rate with asymmetric time-phase transmission probabilities and low quantum bit error rate in the phase basis at 4~dB channel loss.}
		\label{Experimental_Results2}
		\end{center}
\end{figure}

To achieve higher secret key rates, we perform another set of experiments with $p_\textsf{T} = 0.90$ and $p_\textsf{F} = 0.10$. Additionally, we tune the interference visibility by carefully matching the polarization, temporal and spectral overlap to achieve $e_\textsf{F}<$ 0.030. Figure~\ref{Experimental_Results2} shows the secret key rates as a function of dimension for a 4~dB channel loss. The maximum achievable secret key rate when all $d$ or $d-1$ states are transmitted in the phase basis are shown as black squares, and the secret key rate achieved with just 1 monitoring basis state is shown with blue pentagrams. As before, when all $d$ or $d-1$ states are transmitted, we observe that the theoretical limit maximizes at $d = 8$ and rolls off all the way up to $d = 32$. For the case when only one state is transmitted, the upper bound on the phase error rate is higher as expected, but remains below $0.171$. Unlike the previous cases, here the secret key rate increases for both $d = 4$ and $d = 8$ (blue pentagrams in Fig.~\ref{Experimental_Results2}), indicating that it is possible to achieve high secret-key-rate even when transmitting just one monitoring basis state if the error rate is low enough. For $d = 8$, using just one transmitting state, we achieve $\sim 60\%$ of the secret rate that can be achieved by transmitting all 8 phase states. Additionally, with the $d = 8$ states, we can achieve $\sim 164\%$ of the secret key rates achieved with $d = 2$.

An interpretation of our results is that it is possible to achieve on par or better secret key rates than $d = 2$ using high-dimensional encoding, a low repetition rate transmitter, and low-detection-rate single-photon detectors if a QKD system can generate and detect quantum states with low error rates. Additionally, our results conclusively demonstrate that high-dimensional encoding is efficient (more bits per photon) if the quantum channel loss is low. An advantage of using this protocol is that the dimension of the system can be tuned using only a software without changing anything in the hardware platform. This means that the secret key rate of the QKD system can be characterized rapidly based on the HOM visibility, and the dimension that maximizes the secret key rate can be determined without changing anything in the hardware setup. Such flexibility of the protocol is highly desired, especially in practical field implementations where maximizing secret key rate is the primary goal. This also allows the protocol to scale beyond $d = 2$ with no additional specialty measurement device. 

\section{Conclusion}\label{Conclusion}
There are a few possible directions for extending this work in the future. For example, one possibility to implement the QKD system with just two detectors, without using the one in the time-basis measurement. Specifically, if the BS and the detector in the time-basis measurement are removed, and the basis choice is determined actively by whether or not Bob randomly transmits a phase basis state, is it possible to implement the protocol with just the two detectors? Such a protocol will require a theoretical investigation of the squashing model~\cite{LoSquashing}, which is important for schemes using threshold single-photon counting detectors. Another possibility is to use multi-photon-resolving single-photon detectors for the phase basis measurement, which are now available with high-detection efficiency and low timing-jitter~\cite{Clinton2017}, both of which are crucial for this protocol. It is also of interest to investigate a protocol where an untrusted third-party (Charlie) performs the measurement in an arrangement similar to the twin-field QKD protocol~\cite{Lucamarini2018}. Furthermore, our approach can also be implemented in other high-dimensional QKD protocols that use spatial degrees-of-freedom to encode information~\cite{ding2017high,Sit:17,Cozzolino_2018}. Finally, the infinite-key results that we present here can be promoted to the finite key results using entropic uncertainty principles~\cite{Boyd15}.

\textbf{Acknowledgments}
We gratefully acknowledge the discussion of this work with Norbert L\"utkenhaus. We acknowledge the financial support of the Office of Naval Research Multidisciplinary University Research Initiative program on Wavelength-Agile QKD in a Marine Environment (grant N00014-13-1-0627). C.C.W.L acknowledges support by the National Research Foundation (NRF) Singapore, under its NRF Fellowship programme (NRFF11-2019-0001) and Quantum Engineering Programme (QEP-P2), the National University of Singapore, and the Asian Office of Aerospace Research and Development.

\begin{table*}
\centering
\caption{\textbf{Experimental Parameters.} The mean-photon numbers $\mu_1,~\mu_2$ and $\mu_3$, quantum bit error rates in the time-basis $e_{\textsf{T}, \mu_1}$, $e_{\textsf{T}, \mu_2}$ corresponding to mean-photon numbers $\mu_1$ and $\mu_2$, respectively, two-photon quantum bit error rate in the phase basis $(e_\textsf{F})$, and upper bound on the phase error rate $e_{\textsf{F}}^{\mathsf{U}}$ derived from the SDP analysis are listed as a function of the quantum channel loss for different dimensions ($d$) and transmission probabilities $p_\textsf{T} : p_\textsf{F}$. The horizontal lines in the last column indicate the phase error rates that could not be calculated due to the outer product of the Hilbert space ($d \otimes d$) exceeding the matrix size that Matlab can handle.}
\label{tab:Experimental_Parameters}
\begin{tabular}{|c|c|c|c|c|c|c|c|c|c|} 
\hline
$\text{Loss} = -10\log(\eta_\textsf{ch})~\text{(dB)}$ &~~~~$~~p_\textsf{T} : p_\textsf{F}$~~~~&~~~~$d$~~~~&~~~~ $\mu_1$~~~~&~~~~$\mu_2$~~~~&~~~~$\mu_3$~~~~&~~~~$e_{\textsf{T}, \mu_1}$  ~~~~&~~~~$e_{\textsf{T}, \mu_2}$~~~~&~~~~$e_\textsf{F}$~~~~&~~~~ $e_{\textsf{F}}^{\mathsf{U}}$~~~~\\
\hline
4~   & 0.90:0.10 & \begin{tabular}[c]{@{}l@{}}2\\4\\8\\16\\32\end{tabular} & 0.583  & 0.160 & 0.011 & \begin{tabular}[c]{@{}l@{}}0.010\\0.005\\0.014\\0.016\\0.021\end{tabular} & \begin{tabular}[c]{@{}l@{}}0.027\\0.029\\0.038\\0.063\\0.098\end{tabular} & \begin{tabular}[c]{@{}l@{}}0.015\\0.027\\0.021\\0.030\\0.029\end{tabular} & \begin{tabular}[c]{@{}l@{}}0.015\\0.130\\0.171\\--------\\--------\end{tabular}  \\ 
\hline
4    & 0.50:0.50 & \begin{tabular}[c]{@{}l@{}}2\\4\\8\\16\end{tabular}     & 0.156 & 0.059 & 0.007 & \begin{tabular}[c]{@{}l@{}}0.013\\0.022\\0.022\\0.018\end{tabular}        & \begin{tabular}[c]{@{}l@{}}0.037\\0.040\\0.045\\0.041\end{tabular}        & \begin{tabular}[c]{@{}l@{}}0.058\\0.042\\0.041\\0.035\end{tabular}        & \begin{tabular}[c]{@{}l@{}}0.058\\0.205\\0.328\\--------\end{tabular}     \\ 
\hline
8    & 0.50:0.50 & \begin{tabular}[c]{@{}l@{}}2\\4\\8\\16\end{tabular}     & 0.195 & 0.064 & 0.006 & \begin{tabular}[c]{@{}l@{}}0.017\\0.013\\0.010\\0.018\end{tabular}        & \begin{tabular}[c]{@{}l@{}}0.036\\0.031\\0.022\\0.029\end{tabular}        & \begin{tabular}[c]{@{}l@{}}0.041\\0.037\\0.038\\0.034\end{tabular}        & \begin{tabular}[c]{@{}l@{}}0.041\\0.181\\0.299\\--------\end{tabular}     \\
\hline 
\hline
\end{tabular}
\end{table*}

\bibliography{sample}

\end{document}